\documentclass[10pts,showpacs,preprint,aps]{revtex4}
\usepackage[pdftex] {graphics, graphicx,color}
\usepackage{graphicx}
\usepackage{dcolumn}
\usepackage{bm}
\usepackage{amssymb}
\usepackage{amsmath}
\begin{document}
\setcounter{page}{1}
\vskip 2cm
\title 
{The Planck length as a duality of the Cosmological Constant: S-dS and S-AdS thermodynamics from a single expression}
\author
{Ivan Arraut}
\affiliation{  
Department of Physics, Osaka University, Toyonaka, Osaka 560-0043, Japan}

\begin{abstract}
In this paper we suggest that the Planck length $l_{pl}$ and the Cosmological Constant scale $r_\Lambda=\frac{1}{\sqrt{\Lambda}}$ could in principle be dual each other if we take seriously the so-called q-Bargmann Fock space representation as has been previously suggested by Kempf and others and if additionally we introduce $l_{pl}$ as an ultraviolet cut-off and $r_\Lambda=\frac{1}{\sqrt{\Lambda}}$ as an infrared one. As a consequence, it is possible to demonstrate that a Generalized Uncertainty Principle (GUP) given by $\Delta X \Delta P\geq \frac{\hbar}{2}+\frac{l_{pl}^2}{2\hbar}(\Delta P)^2+\frac{\hbar}{2r_\Lambda^2}(\Delta X)^2$, can reproduce appropriately the thermodynamic behavior for both, the Schwarzschild Anti de-Sitter (S-AdS) and the Schwarzschild de-Sitter (S-dS) space without making any analytic extension for the coefficient (parameter) related to the minimum uncertainty in momentum (already suggested in the literature). This is possible if the Black Hole temperature is described with respect to the "natural" Static Observer for the S-dS case located at a distance $l_0=\left(\frac{3}{2}r_s r_\Lambda^2\right)^{1/3}$.         
\end{abstract}
\preprint{OU-HET 761-2012}
\maketitle 
\section{Introduction}
Every approach to Quantum Gravity agrees in the fact that at the Planck scale ($l_{pl}$) the structure of the spacetime is discrete \cite{W,Mead,B,C, Arraut2, Arraut3}. The Generalized Uncertainty Principle (GUP) has been introduced as a way to understand partially what happens at that scale. It has been used for the possible modifications of the physics for systems such as the Harmonic Oscillator \cite{Lay} and Black Hole Thermodynamics \cite{3}. 
Most of the different analysis are performed by introducing an ultraviolet (UV) cut-off at the Planck scale \cite{Lay}. Other authors also include an infrared (IR) cut-off by using the so-called q-Bargmann Fock formalism \cite{3,4,5}. In such a case, the UV scale is in some sense dual to the IR one. From this formalism, GUP is written as $\Delta X \Delta P\geq 1+l_{pl}^2(\Delta P)^2+\frac{1}{r_\Lambda^2}(\Delta X)^2$ ($\hbar=1$) if we impose the $l_{pl}$ as UV cut-off and the Cosmological Constant $\Lambda=\frac{1}{r_\Lambda^2}$ as an IR one. 
The introduction of $\Lambda$ in the formalism of GUP \cite{Arraut, Brett, Arraut4} has already been studied in the literature. In \cite{Arraut}, the derivation was heuristic and following the analogies suggested in \cite{3}. In such a case, GUP was written with $\Delta X$ as a function of $\Delta P$ and assuming the validity of the condition $\Delta P\approx \frac{1}{\Delta X}$, which is related to the negative heat capacity for Black Holes \cite{3, Arraut, Arraut4}. Under this approximation we observe that the proposal given in \cite{Arraut} is just equivalent to that suggested in \cite{Brett} for the S-dS case if the parameter related to the IR cut-off is said to be analytically continued into the imaginary plane. 
There are however some inconsistences if we simply make an analytic continuation. In \cite{Brett} for example, as $\Delta X>>l_{pl}$, GUP was written for the S-dS case as $\Delta P\geq\frac{1}{\Delta X}-\frac{1}{r_\Lambda^2}\Delta X$. Although $\Lambda$ is supposed to be introduced as an IR cut-off, it is clear that the previous expression simply gives $\Delta P_{min}=0$ as $\Delta X=r_\Lambda$. But in Quantum Mechanics language this means that it would be possible to define exactly the momentum of a particle and still have a finite uncertainty in position. Additionally, if we accept the analogy suggested by Adler and Santiago \cite{3}, namely $\Delta P\approx \kappa$ (the surface gravity) and $\Delta X\approx r_H$ (the event horizon); then the result obtained in \cite{Brett} is clearly in contradiction with the minimum temperature found by Bousso and Hawking for the S-dS Black Hole thermodynamics \cite{2}. 
On the other hand, the approach followed in \cite{Arraut} has the problem that it was assumed that $\Delta X=r_H\approx 2GM$. But the event horizon for the S-dS black hole is not exactly given by $2GM$, this is just an approximation valid as the Black Hole mass satisfies the condition $M<<M_{max}=\frac{1}{3}\frac{m_{pl}^2}{m_\Lambda}$ \cite{Arraut}. Additionally, it was not clear how to fix the parameter related to the IR cut-off and the value taken by that parameter fix the minimum temperature. In \cite{Arraut} for example, GUP is written as $\Delta X\geq \frac{1}{2\Delta P}-\frac{\gamma}{3}\frac{m_\Lambda^2}{(\Delta P)^3}$, where $r_\Lambda=\frac{1}{m_\Lambda}$ and it was used the validity of the approximation $\Delta X\approx \frac{1}{\Delta P}$ (negative heat capacity condition for BH). In such a case, GUP could be written as $\Delta P\geq \frac{1}{2\Delta X}-\frac{\gamma}{3}\frac{m_\Lambda^2}{(\Delta P)^2\Delta X}$ or $\Delta P\geq \frac{1}{2\Delta X}-\frac{\gamma'}{3}m_\Lambda^2(\Delta X)$. If for example $\gamma'=\frac{3}{2}$, then $\Delta P_{min}=0$ as $\Delta X_{max}=r_\Lambda$. We have then some degree of arbitrariness for the GUP expression suggested in \cite{Arraut}.\\
In the case of S-AdS, it is already known that there is also a minimum temperature for the Black Hole \cite{Hawking}. This temparature is obtained as the Black Hole has a mass given by $M_{crit}=\frac{2}{3}\frac{m_{pl}^2}{m_\Lambda}$. If the mass is bigger or smaller than this value, then the Black Hole temperature increases. Then the S-AdS Black Hole has a negative heat capacity as $M<M_{crit}$ and a positive heat capacity one as $M>M_{crit}$. 
It is trivial to show that if $\Delta X=r_+$ (here we will denote $r_+$ as the event horizon for the S-AdS space) and $\Delta P=\kappa$, GUP with UV and IR cut-offs, reproduces the appropriate results for S-AdS without any problem or contradiction. This is possible since GUP given by $\Delta X \Delta P\geq 1+l_{pl}^2(\Delta P)^2+\frac{1}{r_\Lambda^2}(\Delta X)^2$ has four solutions. Two of them complement providing the negative heat capacity region for the Black Hole temperature with a minimum and maximum temperature. This portion agrees for both, de-Sitter and Anti de-Sitter spaces. The other two solutions represent the positive heat capacity region for the AdS black hole and the transplanckian (modes) temperatures respectively. For the AdS case, the event horizon increases with the mass \cite{Hawking}. The two solutions with negative Heat Capacity are joined at the UV-IR mix scale, defined as the geometric average of the $l_{pl}$ scale and the $\Lambda$ one ($l_0=(l_{pl}r_\Lambda)^{1/2}$). This scale represents an extremal condition for GUP. At this scale the uncertainty principle (in generalized form) takes its minimum value.\\   
In this paper we demonstrate that GUP with UV and IR cut-offs, can in principle explain in a satisfactory way the Thermodynamic associated to the de-Sitter space without making any analytic continuation (changing the sign of $\Lambda$) for the parameter related to the minimum uncertainty in momentum. This is possible if the temperature for the S-dS case is defined with respect to the "natural" static observer position $\left(r_g=\left(\frac{3}{2}r_s r_\Lambda^2\right)^{1/3}\right)$ defined as the distance at which the attractive effects due to gravity and the repulsive ones due to $\Lambda$ just cancel.\\
The paper is organized as follows: In Sections \ref{eq:dS BH} and \ref{eq:s3}, we make a brief review for the standard de-Sitter Black Hole thermodynamics as has been already performed by Bousso and Hawking \cite{2}. In section \ref{eq:GUP}, we make a brief review of GUP with UV cut-off as has been already performed by Adler, Santiago and others. In section \ref{eq:UV-IR} we introduce the formalism developed by Kempf and we suggest the $l_{pl}$ scale as an UV cut-off and the $\Lambda$ scale as an IR one, we then make the formal derivation of the IR-UV mix scale given given by $l_0=(l_{pl}r_\Lambda)^{1/2}$ as an extremal condition for GUP. In section \ref{eq:MSPM}, we solve GUP in order to obtain the different solutions of the quadratic GUP equation. In section \ref{eq:I}, we explain how can be obtained the de-Sitter Black Hole temperature when the observations are defined by the "natural" Static Observer $r_g$. In Section \ref{eq:Ads}, we make a brief review for the AdS thermodynamic already performed by Hawking and Page \cite{Hawking} and we explain how the different solutions for GUP obtained in section \ref{eq:MSPM} are related to it. In section \ref{eq:Remds}, we make a comparison between the dS Black Hole temperature and the AdS one and we explain the similarities when the dS thermodynamic is analyzed with respect to the "natural" Static Observer. And finally, in section \ref{eq:Conc} we conclude.
     
\section{Horizons in a de-Sitter black hole}   \label{eq:dS BH}

As a starting point, we will first derive the event horizons corresponding to the Black Hole in an asymptotically de-Sitter space. The Schwarzschild de-Sitter (SD) metric is given by \cite{1}: 

\begin{equation}   \label{eq:4}
ds^2=-V(r)dt^2+V(r)^{-1}dr^2+r^2d\Omega_2^2
\end{equation}

where:

\begin{equation}   \label{eq:5}
V(r)=1-\frac{r_s}{r}-\frac{1}{3}\frac{r^2}{r_\Lambda^2}\;\;\;\;\;r_s=2G_N M\;\;\;\;\;r_\Lambda=\frac{1}{\sqrt{\Lambda}}
\end{equation}

the event horizons for this metric can be calculated by using the standard condition \cite{1}:

\begin{equation}   \label{eq:51}
g^{rr}(r_c)=0
\end{equation} 

the two event horizons corresponding to this space are given by \cite{1}:

\begin{equation}  \label{eq:7} 
r_{CH}=-2r_\varLambda cos\left(\frac{1}{3}\left(cos^{-1}\left(\frac{3r_s}{2r_\varLambda}\right)+2\pi\right)\right)
\end{equation}

and:

\begin{equation*}
r_{BH}=-2r_\varLambda cos\left(\frac{1}{3}\left(cos^{-1}\left(\frac{3r_s}{2r_\varLambda}\right)+4\pi\right)\right)
\end{equation*}

where $r_{CH}$ corresponds to the Cosmological Horizon and $r_{BH}$ corresponds to the Black Hole event horizon. The equations \ref{eq:7} show that the maximum mass for a Black Hole in an universe with a positive Cosmological Constant $\Lambda$ is given by:

\begin{equation}   \label{eq:8}
M_{max}=\frac{1}{3}\frac{m_{pl}^2}{m_\varLambda}
\end{equation}

where $m_{pl}$ corresponds to the Planck mass and $m_\Lambda=\sqrt{\Lambda}$. If the mass of a Black Hole is bigger than the value \ref{eq:8}, then there is no radiation process at all and we only have a naked singularity. $M_{max}$ is however of the order of magnitude of the mass of the Universe if we take the observed value for $\Lambda$ \cite{Arraut}. As $M=M_{max}$, the two event horizons take the same value $\left(r_{BH}=r_{CH}=r_\Lambda=\frac{1}{\sqrt{\Lambda}}\right)$, they are degenerate and there is no net radiation  due to the thermodynamic equilibrium established. For degenerate horizons, the Schwarzschild-like coordinates given by the expression \ref{eq:4} and \ref{eq:5} are not valid anymore \cite{2,222}. 
As Bousso and Hawking have explained before \cite{2,222}, as $M\to M_{max}$, then $V(r)\to0$ between the two horizons (BH and Cosmological). In such a case we need a new coordinate system. In agreement with Ginsparg and Perry \cite{GP}, we can write:

\begin{equation}   \label{eq:9}
9G^2M^2\Lambda=1-3\epsilon^2   \;\;\;\;\;0\leq\epsilon\ll1
\end{equation} 

Where the degenerate case (where the two horizons become the same) corresponds to $\epsilon\to0$. We must then define the new radial and the new time coordinates to be:

\begin{equation}   \label{eq:10}
\tau=\frac{1}{\epsilon\sqrt{\Lambda}}\psi\;\;\;\;\; r=\frac{1}{\sqrt{\Lambda}}\left(1-\epsilon cos\chi-\frac{1}{6}\epsilon^2\right)
\end{equation}

in these coordinates, the Black Hole horizon corresponds to $\chi=0$ and the Cosmological horizon to $\chi=\pi$ \cite{2,222}. The metric then becomes:

\begin{equation}   \label{eq:11}
ds^2=-r_\Lambda^2\left(1+\frac{2}{3}\epsilon cos \chi\right)sin^2\chi d\psi^2+r_\Lambda^2\left(1-\frac{2}{3}\epsilon cos\chi\right)d\chi^2+r_\Lambda^2(1-2\epsilon cos\chi)d\Omega_2^2
\end{equation}  

this metric has been expanded up to first order in $\epsilon$. Eq. \ref{eq:11} is of course the appropriate metric to be used as the mass of the Black Hole is near to its maximum value given by \ref{eq:8}. The importance of the previous analysis will become clear in the next section.

\section{The minimum temperature for a Black Hole in a Schwarzschild-de Sitter (SD) space}   \label{eq:s3}   

In general, the surface gravity, can be calculated as \cite{2,222}:

\begin{equation}   \label{eq:12}
\kappa_{BH, CH}=\left(\frac{(K^\mu\nabla_\mu K_\gamma)(K^\alpha\nabla_\alpha K^\gamma)}{-K^2}\right)^{1/2}_{r=r_{BH, CH}}
\end{equation}

The subscripts BH and CH correspond to the Black Hole and Cosmological Horizons respectively. The spacetime described by the metric of the previous section, admits a timelike Killing vector field. It is given by:

\begin{equation}   \label{eq:13}
K=\gamma_t\frac{\partial}{\partial t}
\end{equation}

the surface gravity in general depends on the selected normalization for the Killing vector field. In other words, it depends on the choice of $\gamma_t$. This is a consequence of the observer dependent notion of the Black Hole temperature\cite{Barcelo} which is related to the static observer position. In an asymptotically flat spacetime, the "natural" static observer is at infinity. In such a case, the normalization of the Killing vector is simply $K^2=-1$ ($\gamma_t=1$) at infinity. 
This normalization cannot be used in general when the spacetime is not asymptotically flat. In the case of an asymptotically de-Sitter spacetime for example, there exist a Cosmological Horizon of the order of magnitude of $r_\Lambda$. In this spacetime (S-dS), $\gamma_t\approx 1$ only as the condition $r_s<<r_\Lambda$ is satisfied. \\
We will define the "natural" static observer position as $r_g$. At this distance, the orbit of the Killing vector coincides with the geodesic through $r_g$ at constant angular variables \cite{2, 222}. In other words, we will take it to be the $S^2$ sphere at which the attractive effects due to gravitation are cancelled exactly with the repulsive ones produced by $\Lambda$.\\
An observer at $r_g$ does not feel any acceleration, he is in an unstable equilibrium. This scale is obtained from the condition \cite{1}:

\begin{equation}   \label{eq:14}
\frac{dg_{00}}{dr}=0
\end{equation}
   
this condition for the S-dS metric gives the result:

\begin{equation}   \label{eq:15}
r_g=\left(\frac{3}{2}r_sr_\Lambda^2\right)^{1/3}
\end{equation} 

with this position for the static observer, the normalization factor for the Killing vector becomes \cite{2, 222}:

\begin{equation}   \label{eq:16}
\gamma_t=U(r_g)^{1/2}=\left(1-\left(\frac{3r_s}{2r_\Lambda}\right)^{2/3}\right)^{1/2}
\end{equation}

and the equation for the surface gravity given in \ref{eq:12}, takes the form:

\begin{equation}   \label{eq:17}
\kappa_{CH, BH}=\frac{1}{2\sqrt{U(r_g)}}\left\vert\frac{\partial U}{\partial r}\right\vert_{r=r_BH,CH}
\end{equation}

where for the S-dS metric we get:

\begin{equation}   \label{eq:18}
\kappa_{BH, CH}=\frac{\left\vert\frac{1}{r_H}-\frac{r_H}{r_\Lambda^2}\right\vert}{2\left(1-\left(\frac{3r_s}{2r_\Lambda}\right)^{2/3}\right)^{1/2}}
\end{equation} 
 
here we use $r_H$ for the Black Hole event horizon or for the Cosmological one. The absolute value is necessary if we want to guarantee a positive definite temperature for both, the Black Hole Horizon and the Cosmological one. Physically, it is due to the fact that for the static observer; any object below his position-namely $r<r_g$ will move toward the Black Hole horizon. On the other hand, any object at distances larger that $r_g$, will move toward the Cosmological Horizon.
In the neighborhood of the maximum mass given by \ref{eq:8}, as we have explained before, the Schwarzschild like coordinates become inappropriate. In agreement with Bousso and Hawking, the surface gravity at this regime is given by:

\begin{equation}   \label{eq:188}
\kappa_{CH, BH}=\frac{1}{r_\Lambda}\left(1\mp\frac{2}{3}\epsilon\right)+O(\epsilon^2)
\end{equation}

corresponding to the Black Hole (+ sign) and the Cosmological Horizon (- sign) respectively. $\epsilon$ is already defined in \ref{eq:9}. The minimum temperature for a Black Hole in an asymptotically de-Sitter space is then given by:

\begin{equation}   \label{eq:19}
\kappa_{min}^{BH}=\frac{1}{r_\Lambda}
\end{equation}      

For the Cosmological Horizon, the minimum temperature is obtained as $M=0$ and it is given by $\kappa_{min}^{CH}=\frac{1}{\sqrt{3}}\frac{1}{r_\Lambda}$, which is smaller than the value $\ref{eq:19}$ but of the same order of magnitude. The Cosmological Horizon has a positive Heat Capacity. In other words, it is stable. We must take into account that the Cosmological Horizon decreases with the mass. This statement becomes important at the moment of making an analogy with the Generalized Uncertainty Principle (GUP) as we will explain later in this paper.    

\section{Ultraviolet (UV) Generalized Uncertainty Principle and Black Hole Thermodynamics} \label{eq:GUP}
 
In this section we will briefly summarize the approach to the Black Hole Thermodynamics via GUP. In this case, we will assume that there is no IR cut-off. Only the UV cut-off will be taken into account. In such a case, GUP provides the following expression \cite{3}:

\begin{equation}   \label{eq:20}
\Delta X \Delta P\geq \frac{1}{2}+\frac{l_{pl}^2}{2}\Delta P^2 
\end{equation}

if we take the uncertainty in position as the Black Hole event horizon $\Delta X\approx r_H$ and the momentum uncertainty as the surface gravity $\Delta P \approx \kappa$ \cite{3}, then we obtain:

\begin{equation}   \label{eq:21}
\kappa \approx \frac{1}{2r_H}+l_{pl}^2\frac{\kappa^2}{2r_H}
\end{equation} 
 
this is just a quadratic equation for the surface gravity. If we solve this equation, then we obtain:

\begin{equation}   \label{eq:22}
\kappa_{1,2}=\frac{r_H}{l_{pl}^2}\left(1\mp\sqrt{1-\left(\frac{l_{pl}}{r_H}\right)^2}\right)
\end{equation}  

where the positive sign corresponds to a positive heat capacity solution ($\kappa_2$) and the negative sign corresponds to the negative heat capacity condition for the standard Black Hole ($\kappa_1$). In agreement with the expression \ref{eq:22}, there is a Black Hole remnant which is obtained as the event horizon takes the value $r_H=l_{pl}$. In such a case, the surface gravity takes its maximum (minimum) value given by:

\begin{equation}   \label{eq:23}
\kappa_{max1}=\kappa_{min2}=\frac{1}{l_{pl}}
\end{equation}

on the other hand, as $r_H>>l_{pl}$, equation \ref{eq:22} gives the following results:

\begin{equation}   \label{eq:24}
\kappa_1\approx \frac{1}{2r_H}\;\;\;\;\;\kappa_2\approx \frac{2r_H}{l_{pl}^2}
\end{equation}  

where $\kappa_1$ corresponds to the Schwarzschild Black Hole solution and $\kappa_2$ would correspond to the transplanckian modes for the temperature with a positive heat capacity region which will not be considered in this manuscript. If $r_H\approx 2GM$ at this regime, then it is easy to verify that $\kappa_1$ is just the Schwarzschild Black Hole surface gravity. Take for example the limit $r_\Lambda\to\infty$ in the expression \ref{eq:18}. In such a case, you will just recover the result \ref{eq:24} for $\kappa_1$ again.

\section{Ultraviolet-Infrared (UV-IR) Generalized Uncertainty Principle (GUP) and Black Hole Thermodynamics}   \label{eq:UV-IR}

If we take seriously the formalism developed by Kempf \cite{4, 5} about the minimum Uncertainties in Position and Momenta, then we can write GUP in a more symmetric formulation with respect to position and momentum as:

\begin{equation}   \label{eq:25}
\Delta X \Delta P \geq \frac{\hbar}{2}\left(1+(q^2-1)\left(\frac{(\Delta X)^2}{4L^2}+\frac{(\Delta P)^2}{4K^2}\right)\right)
\end{equation}

here we have assumed $<X>=<P>=0$ \cite{4,5}. In agreement with the formalism developed in \cite{4,5}, the minimum scale in position corresponds to:

\begin{equation}   \label{eq:26}
\Delta X_{min}=L\sqrt{1-q^{-2}}
\end{equation}  

additionally, it is known that the smallest uncertainty in momentum is given by:

\begin{equation}   \label{eq:27}
\Delta P_{min}=K\sqrt{1-q^{-2}}
\end{equation}

K and L must satisfy the additional constraint:

\begin{equation}   \label{eq:28}
KL=\frac{(q^2+1)\hbar}{4}
\end{equation}

this condition suggests that the UV scale is dual to the IR one. This formulation has two free parameters, which we will fix in agreement with the Planck scale $l_{pl}$ and the Cosmological Constant scale $\Lambda=\frac{1}{r_\Lambda^2}$. We want an expression for GUP symmetric with respect to position and momentum given by:

\begin{equation}   \label{eq:29}
\Delta X \Delta P\geq \frac{\hbar}{2}+\frac{l_{pl}^2}{2\hbar}(\Delta P)^2+\frac{\hbar}{2r_\Lambda^2}(\Delta X)^2
\end{equation}

if we want this expression to agree with \ref{eq:25}, the following conditions must be satisfied:

\begin{equation}   \label{eq:30}
l_{pl}=L\sqrt{1-q^{-2}}   
\end{equation}

\begin{equation*}   
\frac{1}{r_\Lambda}=K\sqrt{1-q^{-2}}
\end{equation*}

these expressions together with the condition \ref{eq:28}, automatically fix the value of q. Although q could take different values in order to satisfy the previous conditions, we select the value of q satisfying the additional constraint $q\geq 1$ \cite{4,5, Kempf}. In such a case, q, which is supposed to be related to the gravitational degrees of freedom, satisfies:

\begin{equation}   \label{eq:31}
q\approx 1+\frac{l_{pl}}{r_\Lambda}
\end{equation}

where $l_{pl}\approx 10^{-35}mt$ is the Planck scale and $r_\Lambda\approx 10^{26} mt$ is the Hubble one, then $q\approx 1+10^{-61}$. If $\Lambda\to0$ then $q=1$ and $l_{pl}\to0$ and vice versa. It means that inside this formalism, the Cosmological Constant $\Lambda$ is related to the minimum scale in position. Without a minimum scale, there is no Cosmological Constant and vice versa.
Originally Kempf \cite{4,5} derived the results \ref{eq:26} and \ref{eq:27} by definiting the function:

\begin{equation}   \label{eq:111}
f(\Delta X, \Delta P):=\Delta X \Delta P-\frac{\hbar}{2}\left(1+(q^2-1)\left(\frac{(\Delta X)^2+<X>^2}{4L^2}+\frac{(\Delta P)^2+<P>^2}{4K^2}\right)\right)
\end{equation}  

here, we will assume $<X>=0=<P>$. The minimum scale in position can be found given the following extremal condition:

\begin{equation}   \label{eq:112}
\frac{\partial}{\partial \Delta P}f(\Delta X, \Delta P)=0\;\;\;\;\;f(\Delta X, \Delta P)=0
\end{equation}

then, the result is just the eq. \ref{eq:30} $\left(\Delta X_{min}=L\sqrt{1-q^{-2}}\right)$. On the other hand, the minimum scale in momentum is obtained from the condition:

\begin{equation}   \label{eq:113}
\frac{\partial}{\partial \Delta X}f(\Delta X, \Delta P)=0\;\;\;\;\;f(\Delta X, \Delta P)=0
\end{equation}

the result is just the eq. \ref{eq:30} $\left(\Delta P_{min}=K\sqrt{1-q^{-2}}\right)$. We can however, derive a third scale given by the UV-IR mix effects. This scale was introduced for first time by John A. Wheeler \cite{W} in 1957. It is given by the geometrical average of the $l_{pl}$ and $r_\Lambda$, namely, $l_0=(l_{pl}r_\Lambda)^{1/2}$.
We can define the total differential for the function $f(\Delta X, \Delta P)$ as:

\begin{equation}   \label{eq:114}
df(\Delta X, \Delta P)=\left(\frac{\partial f(\Delta X, \Delta P)}{\partial \Delta P}\right)_{\Delta X=C}d(\Delta P)+\left(\frac{\partial f(\Delta X, \Delta P)}{\partial \Delta X}\right)_{\Delta P=C}d(\Delta X)
\end{equation}

The general extremal condition inside the phase space is obtained as the total differential \ref{eq:114} goes to zero. In such a case:

\begin{equation}   \label{eq:115}
df(\Delta X, \Delta P)=0
\end{equation}

obtaining then the result:

\begin{equation}   \label{eq:116}
\frac{d(\Delta X)}{d(\Delta P)}=-\frac{\Delta X}{\Delta P}\frac{\left(1-\frac{\hbar}{4K^2}(q^2-1)\frac{\Delta P}{\Delta X}\right)}{\left(1-\frac{\hbar}{4L^2}(q^2-1)\frac{\Delta X}{\Delta P}\right)}
\end{equation}

let's impose now the constraint $f(\Delta X, \Delta P)=0$. In such a case \ref{eq:111} becomes:

\begin{equation}   \label{eq:117}
\Delta X \Delta P=\frac{\hbar}{2}\left(1+(q^2-1)\left(\frac{(\Delta X)^2+<X>^2}{4L^2}+\frac{(\Delta P)^2+<P>^2}{4K^2}\right)\right)  
\end{equation}

if we differentiate with respect to $\Delta X$ we have:

\begin{equation}   \label{eq:118}
\frac{d}{d(\Delta X)}(\Delta X\Delta P)=\Delta P+\Delta X\frac{d(\Delta P)}{d (\Delta X)}=\frac{\hbar}{4}\left((q^2-1)\left(\frac{\Delta X}{L^2}+\frac{\Delta P}{K^2}\frac{d(\Delta P)}{d(\Delta X)}\right)\right)  
\end{equation}

this equation is just the condition \ref{eq:115} as the reader can easily verify. The same result is obtained when the differentiation is done with respect to $\Delta P$. In such a case:

\begin{equation}   \label{eq:119}
\frac{d}{d(\Delta P)}(\Delta X\Delta P)=\Delta X+\Delta P \frac{d(\Delta X)}{d (\Delta P)}=\frac{\hbar}{4}\left((q^2-1)\left(\frac{\Delta X}{L^2}\frac{d(\Delta X)}{d(\Delta P)}+\frac{\Delta P}{K^2}\right)\right)
\end{equation} 

which is just equivalent to the results \ref{eq:115} and \ref{eq:118}. In fact, the condition \ref{eq:115} is just equivalent to:

\begin{equation}   \label{eq:120}
2d(\Delta X\Delta P)=\frac{\hbar}{4}d\left((q^2-1)\left(\frac{(\Delta X)^2}{L^2}+\frac{(\Delta P)^2}{K^2}\right)\right)
\end{equation}

then the condition $f(\Delta X, \Delta P)=0$ is just included inside \ref{eq:115}. We want to know what is the extremal condition for the Uncertainty Principle $\Delta X \Delta P$. We want to know when $\frac{d(\Delta X \Delta P)}{d\Delta X}=0=\frac{d(\Delta X \Delta P)}{d\Delta P}$. In such a case, from \ref{eq:118} it is clear that we have to satisfy: 

\begin{equation}   \label{eq:121}
\frac{\Delta X}{\Delta P}=-\frac{d(\Delta X)}{d(\Delta P)}
\end{equation}

introducing this result inside the right-hand side of \ref{eq:118} or \ref{eq:119}, we obtain:

\begin{equation}   \label{eq:122}
(q^2-1)\left(\frac{\Delta X}{L^2}+\frac{\Delta P}{K^2}\frac{d(\Delta P)}{d(\Delta X)}\right)=0
\end{equation}

this equation has two solutions. The first one is not interesting for us, because it suggests $q=1$ which is a trivial condition because in such a case $d(\Delta X \Delta P)=0$ everywhere. Additionally, $q=1$ corresponds to the standard Bosonic algebra in agreement with \cite{1, 4, 5}. We then do not consider that case here. The interesting case is:

\begin{equation}   \label{eq:123}
\left(\frac{\Delta X}{L^2}+\frac{\Delta P}{K^2}\frac{d(\Delta P)}{d(\Delta X)}\right)=0
\end{equation}   

which in combination with \ref{eq:121} gives:

\begin{equation}   \label{eq:124}
\Delta X=\pm \frac{L}{K}\Delta P
\end{equation}

if we compare the expressions \ref{eq:29} with the one obtained in \ref{eq:117} under the condition $f(\Delta X, \Delta P)=0$, then it is simple to verify that ($\hbar=1$):

\begin{equation}   \label{eq:125}
L=K^{-1}=\frac{\sqrt{2}}{2}(l_{pl}r_\Lambda)^{1/2}
\end{equation}

then the condition \ref{eq:124} becomes:

\begin{equation}   \label{eq:126}
\Delta X=\frac{1}{2}(l_{pl}r_\Lambda)\Delta P
\end{equation}

here we have only taken into account the positive sign. For consistence, it is easy to verify that the condition \ref{eq:28} is satisfied. It can be verified if we take into account that in agreement with \ref{eq:31}, we have $q^2\approx 1+2\frac{l_{pl}}{r_\Lambda}$. Then \ref{eq:28} becomes ($\hbar=1$):

\begin{equation}   \label{eq:127}
KL\approx\frac{1}{2}
\end{equation} 

This result is consistent with \ref{eq:125} as can be verified. If we replace \ref{eq:126} inside \ref{eq:29}, we then obtain under the approximation $r_\Lambda>>l_{pl}$, the following result:

\begin{equation}   \label{eq:128}
\Delta X\approx (l_{pl}r_\Lambda)^{1/2}=l_0\;\;\;\;\;\Delta P\approx \frac{1}{(l_{pl}r_\Lambda)^{1/2}}=\frac{1}{l_0}
\end{equation}

The result \ref{eq:128} is just the UV-IR scale already defined in \cite{Arraut2} and obtained for first time by John A. Wheeler in 1957 \cite{W} and interpreted as a coherence region.
   
\section{The minimum scale in position and momentum and Black Hole thermodynamics}   \label{eq:MSPM}

Here we want to show that in principle it is possible to reproduce the S-dS Thermodynamics with an UV cut-off without making any analytic extension for the IR cut-off scale in the GUP expression. In order to find the behavior of $\Delta X$ or $\Delta P$ near their minimum values, we must solve for $\Delta X$ near the $r_\Lambda$ scale and for $\Delta P$ near the $l_{pl}$ scale. The GUP expression solved for $\Delta X$ is valid as $\frac{1}{\sqrt{l_{pl}r_\Lambda}}\geq\Delta P\geq\frac{1}{r_\Lambda}$. On the other hand, the GUP expression solved for $\Delta P$ is valid as $(l_{pl}r_\Lambda)^{1/2}\geq\Delta X\geq l_{pl}$. Both solutions are equal at the UV-IR mix scale given by $\Delta X=l_0=(l_{pl}r_\Lambda)^{1/2}$. 
Note that the GUP equation \ref{eq:29} is quadratic for both, position and momentum. We then expect two solutions of $\Delta X$ and two more for the solutions with respect to $\Delta P$. We want to choose the solutions such that as $\Delta X>>l_{pl}$ and $\Delta P>>\frac{1}{r_\Lambda}$ we can then guarantee a negative heat capacity behavior for the Black Hole event horizon in agreement with the analogy suggested by Chen, Adler and Santiago \cite{3}. In such a case, $\Delta X=r_H$ and $\Delta P=\kappa$ correspond to the event horizon and the surface gravity respectively. One of the lessons learned from the S-dS metric is that we cannot say in this analogy that $\Delta X\approx 2GM$. This is just an approximation valid as $M<<M_{max}$ or $\Delta X<<r_\Lambda$ in agreement with eq. \ref{eq:8} and \cite{Arraut}. For $\hbar=1$, let's solve for $\Delta X=r_H$ in equation \ref{eq:29}. In such a case, we obtain:

\begin{equation}   \label{eq:41}
\Delta X_1\approx\left(r_\Lambda^2\Delta P_1\right)\left(1-\sqrt{1-\frac{1}{(\Delta P_1 r_\Lambda)^2}}\right)
\end{equation}   

as $\Delta P_1>>\frac{1}{r_\Lambda}$, the condition $\Delta X_1\approx \frac{1}{\Delta P_1}$ is satisfied. This is what is needed if we want to satisfy the negative heat capacity condition for Black Holes. On the other hand, in the neighborhood of $l_{pl}$, it is better to solve for $\Delta P$. In such a case, we have:

\begin{equation}   \label{eq:42}
\Delta P_1\approx\left(\frac{\Delta X_1}{l_{pl}^2}\right)\left(1-\sqrt{1-\frac{l_{pl}^2}{(\Delta X_1)^2}}\right)
\end{equation}

as $\Delta X_1>>l_{pl}$, we recover again the standard result $\Delta P_1\approx\frac{1}{\Delta X_1}$. The solutions \ref{eq:41} and \ref{eq:42} both satisfy the ordinary Uncertainty principle as $\Delta P_1>>\frac{1}{r_\Lambda}$ and $\Delta X_1>>l_{pl}$. They describe the same solution at different scale regimes of eq. \ref{eq:29}. Eq. \ref{eq:41} is the solution valid when the condition $\frac{1}{\sqrt{l_{pl}r_\Lambda}}\geq\Delta P_1\geq\frac{1}{r_\Lambda}$ is satisfied. On the other hand, \ref{eq:42} is the solution valid when $(l_{pl}r_\Lambda)^{1/2}\geq\Delta X_1\geq l_{pl}$ is satisfied. 
The scale at which \ref{eq:41} and \ref{eq:42} provide the same result is:

\begin{equation}   \label{eq:45}
\Delta X_1=\sqrt{l_{pl}r_\Lambda}\;\;\;\;\;\Delta P_1=\frac{1}{\sqrt{l_{pl}r_\Lambda}}
\end{equation}

equation \ref{eq:45} shows the scale at which the UV modes are perfectly mixed with the IR ones. We have already derived formally this scale in \ref{eq:128}. There are still two more solutions related to \ref{eq:41} and \ref{eq:42}. They correspond to the positive sign solution for the quadratic equations.  Let's rewrite the complement solutions to \ref{eq:41} and \ref{eq:42}. They are:

\begin{equation}   \label{eq:411}
\Delta X_2=\left(r_\Lambda^2\Delta P_2\right)\left(1+\sqrt{1-\frac{1}{(\Delta P_2 r_\Lambda)^2}}\right)
\end{equation} 

and:

\begin{equation}   \label{eq:422}
\Delta P_2=\left(\frac{\Delta X_2}{l_{pl}^2}\right)\left(1+\sqrt{1-\frac{l_{pl}^2}{(\Delta X_2)^2}}\right)
\end{equation}

as $\Delta P_2>>\frac{1}{r_\Lambda}$, then \ref{eq:411} becomes $\Delta X_2\approx r_\Lambda^2\Delta P_2$. In Black Hole thermodynamics, this corresponds to the positive heat capacity region for the surface gravity in the case of S-AdS Black Hole. We will explain this point in detail in future sections. 
On the other hand, as $\Delta X_2>>l_{pl}$, the solution \ref{eq:422} becomes $\Delta P_2\approx \frac{\Delta X_2}{l_{pl}^2}$. This solution will not be considered here since it is related to the transplanckian modes.

\begin{figure}
	\centering
		\includegraphics[scale=1.0]{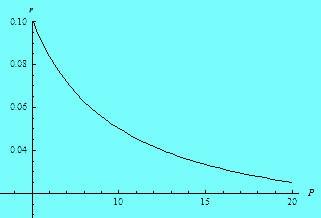}
	\caption{Distance vs Momentum in agreement with the solution \ref{eq:41}.
	$\Lambda$ has been normalized to unity for convenience.}
	\label{fig:GUPlot}
\end{figure}

\begin{figure}
	\centering
		\includegraphics[scale=1.0]{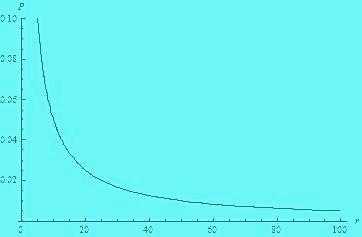}
	\caption{Momentum vs distance in agreement with the solution \ref{eq:42}. $l_{pl}$ has been normalized to unity for convenience. It is evident the negative heat capacity behavior}
	\label{fig:GUPlot2}
\end{figure}

\section{Interpretation}   \label{eq:I} 

From the result \ref{eq:41}, it is clear that there exist a minimum scale in momentum given by $\Delta P_{1min}=\frac{1}{r_\Lambda}$. The corresponding maximum position scale is $\Delta X_1=r_\Lambda$. This is consistent with the results obtained in \ref{eq:7} and \ref{eq:19} in agreement with the S-dS Black Hole as it is described by the "natural" static observer. Then GUP with an infrared cut-off given by $\Lambda$, can describe the same results as in the case of the S-dS Black Hole Thermodynamics related to the Black Hole event Horizon as far as $r_H>> l_{pl}$. We identify $\Delta X_1=r_H$ and $\Delta P_1=\kappa$. In Fig. \ref{fig:GUPlot} we can see the plot of $\Delta X_1=r$ vs $\Delta P_1=P$ for the regime $(l_{pl}r_\Lambda)^{1/2}\leq\Delta X_1\leq r_\Lambda$ or $\frac{1}{\sqrt{l_{pl}r_\Lambda}}\geq\Delta P_1\geq\frac{1}{r_\Lambda}$. Fig \ref{fig:GUPlot2} corresponds to eq. \ref{eq:42}. It is just the continuation of the solution \ref{eq:41} but in the regime $l_{pl}\leq\Delta X_1\leq(l_{pl}r_\Lambda)^{1/2}$ or $\frac{1}{l_{pl}}\geq\Delta P_1\geq \frac{1}{(l_{pl}r_\Lambda)^{1/2}}$. Near the Planck scale the behavior deviates from the standard Black Holes in S-dS space. 
The S-dS thermodynamics is incomplete if we cannot reproduce the surface gravity for the Cosmological Horizon. 
The surface gravity in S-dS metric can be described with respect to the "natural" static observer in agreement with Bousso and Hawking \cite{Hawking}. In GUP language, this is equivalent to a redefinition of the UV and IR cut-offs for the Cosmological Horizon. If the "natural" static observer defines the same origin of coordinates for the Black Hole event Horizon and the Cosmological one, then he must redefine the cut-offs in order to make a description about the surface gravity for the Cosmological Horizon.  
If the BH event horizon is $r_{BH}^0=l_{pl}$, then the Cosmological Horizon is approximately $r_{CH}^0\approx \sqrt{3}r_\Lambda$ and the initial position for the Static observer is $r_{0g}=\left(\frac{3}{2}l_{pl}r_\Lambda^2\right)^{1/3}$.\\
In agreement with the "natural" static observer, the IR cut-off for the Cosmological Horizon is given by $\Delta X_1^0\approx \sqrt{3}r_\Lambda$, which corresponds to $\Delta P_{1min}^{CH}\approx\frac{1}{\sqrt{3}}\frac{1}{r_\Lambda}$. This is just the minimum value of the surface gravity for the Cosmological Horizon already found in section \ref{eq:s3} for the standard S-dS Black Hole. If the static observer defines a different IR scale for the Cosmological Horizon, he must also define a new UV scale. The new UV for the static observer point of view is given by $\Delta X_{1min}^{CH}\approx r_\Lambda$ corresponding to $\Delta P_{1max}^{CH}=\frac{1}{r_\Lambda}$. This conditions and the appropriate behavior for the Cosmological Horizon surface gravity can be obtained from the equations \ref{eq:41} and \ref{eq:42} if we replace $r_\Lambda\to\sqrt{3}r_\Lambda$ and $l_{pl}\to r_\Lambda$. In such a case, the corresponding equations describing the Cosmological Horizon surface gravity are:

\begin{equation}   \label{eq:4111}
\Delta X_1\approx\left(3r_\Lambda^2\Delta P_1\right)\left(1-\sqrt{1-\frac{1}{(\Delta P_1 \sqrt{3}r_\Lambda)^2}}\right)
\end{equation}   

and

\begin{equation}   \label{eq:4222}
\Delta P_1\approx\left(\frac{\Delta X_1}{r_\Lambda^2}\right)\left(1-\sqrt{1-\frac{r_\Lambda^2}{(\Delta X_1)^2}}\right)
\end{equation}

again the figures \ref{fig:GUPlot} and \ref{fig:GUPlot2} under the appropriate normalizations for the cut-offs, represent the behavior for the Cosmological Horizon surface gravity if we take into account that $r_{CH}$ decreases as the mass of the Black Hole increases. It means that the Cosmological Horizon has a positive heat capacity and is stable. If \ref{eq:4111} and \ref{eq:4222} represent extensions of the same graphic in different regimes, it is again clear that there must be a point at which both became the same. This point would be the UV-IR mix scale for the Cosmological Horizon in agreement with the observations of the "natural" static observer. This new scale is obtained by introducing the condition $\Delta X_1=\sqrt{3}r_\Lambda^2\Delta P_1$ inside the GUP expression given in \ref{eq:29}. After replacing this condition in \ref{eq:4111} and \ref{eq:4222}, it is trivial to find that the UV-IR scale for the Cosmological Horizon is also of the order of magnitude of $r_\Lambda$.
It seems then that the S-dS surface gravities corresponding to both, the Black Hole event horizon and the Cosmological one, can be described by the the solutions \ref{eq:41} and \ref{eq:42} if we redefine the UV and IR cut-offs for the Cosmological Horizon case. The solutions \ref{eq:411} and \ref{eq:422} seem to be unnecessary for the description of the S-dS Black Hole from the "natural" static observer point of view.

\section{The Anti-de Sitter Black Hole temperature and GUP}   \label{eq:Ads} 

\begin{figure}
	\centering
		\includegraphics[scale=1.0]{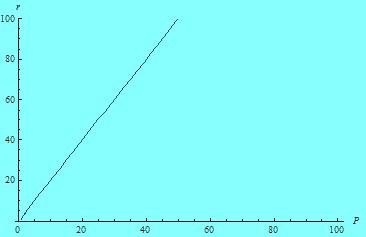}
	\caption{Distance vs Momentum in agreement with the solution \ref{eq:411}. $r_\Lambda$ has been normalized to unity for convenience. It is evident the positive heat capacity behavior if the event horizon $r_+$ increases with the mass.}
	\label{fig:GUP4}
\end{figure}

The Anti-de Sitter space is characterized by a metric written as in eq. \ref{eq:4} but with an opposite sign for $\Lambda$. In \cite{Hawking}, the Anti-de Sitter Black Hole temperature is given by:

\begin{equation}   \label{eq:46}
T=\frac{1}{4\pi}\frac{r_\Lambda^2+r_+^2}{r_\Lambda^2r_+}
\end{equation}

$r_+$ is the event horizon for the A-dS Black Hole. In this case, it is trivial to demonstrate that there is only one event horizon which increases with the mass. It is approximately $r_+\approx 2GM$ as the condition $M<<M_{crit}$ is satisfied. $M_{crit}$ is the mass at the Black Hole temperature takes a minimum value given by $T_{min}=\frac{1}{2\pi}\frac{1}{r_\Lambda}$ \cite{Hawking}. This is in fact consistent with the result \ref{eq:41} as $(l_{pl}r_\Lambda)^{1/2}\leq\Delta X_1\leq r_\Lambda$. In \ref{eq:46}, we assume $r_\Lambda=\frac{1}{\sqrt{-\Lambda}}$ with $\Lambda<0$.
The relation between the event horizon and the mass of the A-dS Black Hole is given by \cite{Hawking}:

\begin{equation}   \label{eq:47}
M=\frac{1}{2}m_{pl}^2r_+\left(1+\frac{r_+}{3r_\Lambda}\right)
\end{equation}

$M_{crit}$ can be found as the mass after which the Black Hole temperature increases with the mass, it is simply:

\begin{equation}   \label{eq:48}
M_{crit}=\frac{2}{3}\frac{m_{pl}^2}{m_\Lambda}
\end{equation}
  
as $M=M_{crit}$, then $r_+=r_\Lambda=\frac{1}{m_\Lambda}$. As $M>M_{crit}$, the temperature increases with the mass (positive heat capacity). As $M<M_{crit}$, the temperature decreases with the increasing mass (negative heat capacity). It is clear from \ref{eq:46} that as $r_+<<r_\Lambda$, $T \approx \frac{1}{4\pi r_+}$ which is just the standard Schwarzschild behavior. \\
The expressions \ref{eq:41} and \ref{eq:42} obtained from GUP describe the appropriate behavior for the A-dS Black Hole as far as $M<M_{max}$ (see the figures \ref{fig:GUPlot} and \ref{fig:GUPlot2}), this corresponds to the negative heat capacity region.\\
On the other hand, as $M>M_{max}$ the temperature increases with the mass. If $r_+>>r_\Lambda$, then $T\approx \frac{1}{4\pi}\frac{r_+}{r_\Lambda^2}$. If we again accept the analogy $\Delta X_2\approx r_+$ and $\Delta P_2\approx \kappa=2\pi T$, then the equation \ref{eq:411} which is one of the solutions for GUP, can describe the appropriate behavior for the A-dS black Hole as $M>M_{crit}$. Remember that we have already demonstrated that as $\Delta P_2>>\frac{1}{r_\Lambda}$, the GUP solution \ref{eq:411} is approximately $\Delta P_2\approx\frac{\Delta X_2}{r_\Lambda^2}$ which has the same linear behavior for A-dS Black Hole in the same regime (see the figure \ref{fig:GUP4}).

\section{A deep comparison between the S-dS and S-AdS Black Hole temperatures}   \label{eq:Remds}

It is interesting to note that the thermodynamic behavior of a S-dS Black Hole is just similar to that obtained for the A-dS one as $M\leq M_{crit}$ if the temperature is measured by the "natural" Static Observer. In both cases, we have a negative heat capacity and the same minimum temperature $\kappa=\sqrt{\vert\Lambda\vert}$. We also have the same value for the event horizon as the Holes reach the minimum temperature, they are $r_{BH}=r_+=r_\Lambda$ as $T\to T_{min}\approx \frac{1}{r_\Lambda}$. This is not a coincidence since both spaces have the same IR cut-off scale although different geometry and causal structure. As far as $M<<M_{max}$ and also $M<<M_{crit}$, both, the S-dS Black Hole event horizon and the A-dS one, are approximately $r_{BH}\approx 2GM\approx r_+$. In such a case, the conditions $r_{BH}<<r_\Lambda$ and $r_+<<r_\Lambda$ are also satisfied \cite{Arraut}. Then the Killing vector normalization in agreement with the "natural" Static Observer position (see equation \ref{eq:16}) is just $K^2\approx -1$ for the S-dS case (see eq. \ref{eq:18}) and:

\begin{equation}   \label{eq:49}
\kappa_{BH}=\kappa_{AdS}=\frac{1}{r_{BH,+}}
\end{equation}      

the GUP modifications near the UV regime applies for both cases, S-dS and S-AdS (see eq. \ref{eq:42}). If $M<<M_{max}$ we can make an expansion for eq. \ref{eq:18}. in order to obtain the first order corrections as:

\begin{equation}   \label{eq:52}
\kappa_{dS}\approx \frac{1}{2}\left(1+\frac{1}{2}\left(\frac{3}{2}\frac{r_s}{r_\Lambda}\right)^{2/3}\right)\frac{1}{r_{BH}}
\end{equation}

the first order correction is due to the normalization of the Killing vector and the fact that the temperature is measured by the "natural" static observer. This correction increases the effective coefficient of the term proportional to $\frac{1}{r_H}$. In this way, the temperature still decreases with the increasing mass (negative heat capacity) but at a smaller rate in comparison with the standard surface gravity with a Killing vector normalization $K^2=-1$.
We can check what happens in the same regime with the A-dS Black Hole temperature. We can make the same expansion for the A-dS black Hole and compare with \ref{eq:52}. If we assume that in the mentioned regime $M<<M_{crit}$ (for the A-dS case), then we can assume $r_+\approx 2GM$ and the temperature \ref{eq:46} would be proportional to:

\begin{equation}   \label{eq:53}
\kappa_{AdS}\approx \frac{1}{2r_+}+\frac{r_+}{2r_\Lambda^2}
\end{equation}    

if we want to compare with \ref{eq:52}, we must use the approximations $r_{BH}\approx 2GM$ as $M<<M_{max}$ for the S-dS black hole and $r_+\approx 2GM$ as $M<<M_{crit}$ for S-AdS. In such a case, the expressions \ref{eq:52} and \ref{eq:53} become:

\begin{equation}   \label{eq:54}
\kappa_{dS}\approx \frac{1}{4GM}+\frac{9^{1/3}}{8}\frac{1}{(r_\Lambda^2 GM)^{1/3}}
\end{equation}

and:

\begin{equation}   \label{eq:55}
\kappa_{AdS}\approx \frac{1}{4GM}+\frac{GM}{r_\Lambda^2}
\end{equation}

the first order corrections obtained in equations \ref{eq:54} and \ref{eq:55} as $M<<M_{max}=\frac{1}{2}M_{crit}$ are in general negligible for most of the mass range for both Black Holes (dS and AdS). 
Now we can expand both temperatures as $r_{BH}\to r_\Lambda$ and $r_+\to r_\Lambda$. We can assume for example that $r_+=Cr_\Lambda$, where C is some arbitrary constant near to the unit value. In such a case, \ref{eq:46} can be written as:

\begin{equation}   \label{eq:56}
\kappa_{AdS}\approx\frac{1}{r_\Lambda}\left(\frac{1}{2C}+\frac{C}{2}\right)
\end{equation}

the expansion for $\kappa_{dS}$ near $\kappa_{min}$ has been already performed by Bousso and Hawking. It is given by \ref{eq:188} and repeated here for clarity:

\begin{equation}   \label{eq:57}
\kappa_{dS}\approx \frac{1}{r_\Lambda}\left(1+\frac{2}{3}\epsilon\right)
\end{equation}
   
if we want the two expressions, \ref{eq:56} and \ref{eq:57} to be the same, it is necessary to satisfy the constraint:

\begin{equation}   \label{eq:58}
C=1+\frac{2}{3}\epsilon\pm\frac{2}{\sqrt{3}}\sqrt{\epsilon}
\end{equation}

the negative sign is the relevant to our comparison. The positive sign solution, corresponds to the positive heat capacity region for the AdS Black Hole. We see that as $M=M_{max}$ for S-dS, then $\epsilon \to0$ and $C=1$, which is completely consistent. The opposite is also true, if $C=1$, then $M=M_{crit}$ for the AdS black hole and then $\epsilon\to0$. Then if the surface gravity for the S-dS Black Hole is defined with respect to the "natural" Static Observer position, then there is a mass region where there is no distinction between the S-dS Black Hole temperature and the AdS one. This fact permits us to obtain a single temperature expression (GUP) for both, de-Sitter and anti-de Sitter space.      

\section{Conclusions}   \label{eq:Conc} 

In this paper, we have explored the consequences of GUP with an UV cut-off and an IR one. We have demonstrated that the same GUP expression can reproduce the thermodynamic for both; Schwarzschild de-Sitter Black Hole and the Schwarzschild Anti de-Sitter one without making any analytic continuation for the IR cut-off scale.
In the case of S-dS, the extension of the analogy with GUP for the Cosmological Horizon temperature is obtained if we redefine the UV and IR cut-offs in agreement with the conditions imposed by the "natural" Static Observer. This implies a redefinition for the minimum uncertainties in position and momentum for the Cosmological Horizon case. \\
For the S-AdS case, the GUP analogy is natural and it makes use all the solutions obtained from the GUP expression (except the transplanckian modes). GUP can reproduce both, the negative Heat capacity region and the other positive heat capacity one.\\
Finally we made a comparison between the dS Black Hole and the AdS one in order to analyze their thermodynamic similarities. This is motivated by the fact that it seems that GUP with UV and IR cut-off provides appropriate expressions for both spaces with different geometries and different causal structures. Why is this the case, is an inquiry that will be analyzed in a future manuscript by using the full machinery of the q-Bargmann Fock algebras.


\newpage
\noindent
\begin{thebibliography}{99}
\bibitem{W} John A. Wheeler {\it On the nature of Quantum Geometrodynamics}, Ann. Phys. {\bf 2}. 604-614 (1957).
\bibitem{Mead} C. A. Mead {\it Possible connection between Gravitation and Fundamental length}, Phys. Rev. {\bf135}, B849 (1964); M. Maggiore {\it A Generalized Uncertainty Principle in Quantum Gravity}, Phys. Lett. {\bf B 304}, 65 (1993); L. J. Garay {\it Quantum Gravity and minimum length}, Int. J. Mod. Phys. {\bf A 10}, 145 (1995).
\bibitem{B} P. Nicolini, {\it Nonlocal and Generalized Uncertainty Principle Black Holes}, arXiv:1202.2102.
\bibitem{C} P. Nicolini, {\it Noncommutative Black Holes, the final appeal to Quantum Gravity: A Review}, Int. J. Mod. Phys. {\bf A24}, (2009), 1229-1308.
\bibitem{Arraut2} I. Arraut, D. Batic and M. Nowakowski, {\it A Noncommutative model for a Mini Black Hole}, Clas. Quant. Grav. {\bf 26} (2009), 245006.
\bibitem{Arraut3} I. Arraut, D. Batic and M. Nowakowski, {\it Maximal Extension of the Schwarzschild Spacetime inspired by Noncommutative Geometry}, JMP {\bf 51} (2010), 022503.
\bibitem{Lay} L. N. Chang, Z. Lewis, D. Minic and T. Takeuchi {\it On the minimal Length Uncertainty Relation and the Foundations of String Theory}, Adv. High Energy Phys. 2011 (2011), 493514.
\bibitem{3} Pisin Chen, {\it Generalized Uncertainty Principle and Dark Matter}, astro-ph/0305025; Pisin Chen and Ronald J. Adler, {\it Black Hole remnants and Dark Matter}, Nucl. Phys. Proc. Suppl. {\bf 124} (2003), 103-106; Ronald. J. Adler, Pisin Chen and David I. Santiago, {\it The Generalized Uncertainty Principle and Black Hole remnants}, Gen. Rel. Grav. {\bf 33} (2001) 2101-2108.
\bibitem{4} Achim Kempf, {\it On Quantum Field Theory with Nonzero Minimal Uncertainties in Positions and Momenta}, J. Math. Phys. {\bf 38} (1997) 1347-1372.
\bibitem{5} Achim Kempf, {\it Quantum Field Theory with Nonzero Minimal Uncertainties in Positions and Momenta}, hep-th/9405067.
\bibitem{Arraut} I. Arraut, D. Batic and M. Nowakowski, {\it Comparing two approaches to Hawking radiation of Schwarzschild-de Sitter Black Holes}, Class.Quant.Grav. {\bf 26} (2009), 125006.
\bibitem{Brett} B. Bolen and M. Cavaglia, {\it (Anti-) de Sitter Black Hole Thermodynamics and the Generalized Uncertainty Principle}, Gen. Rel. Grav. {\bf 37} (2005), 1255. 
\bibitem{Arraut4} M. Nowakowski and I. Arraut, {\it The Minimum and Maximum temperature of Black Body Radiation}, Mod. Phys. Lett. {\bf A 24} (2009), 2133.
\bibitem{2} R. Bousso and S. W. Hawking, {\it Pair creation of black holes during inflation}, Phys. Rev. {\bf D 54} (1996), 6312-6322.
\bibitem{Hawking} S. W. Hawking and D. N. Page, {\it Thermodynamics of Black Holes in Anti-de Sitter Space}, Commun. Math. Phys {\bf 87}, 577, (1983). 
\bibitem{1} S. L. Ba$\dot{z}\acute{n}$ski and V. Ferrari, {\it Analytic Extension of the Schwarzschild-de Sitter Metric}, Il Nuovo Cimento Vol. 91 B, N. 1, 11 Gennaio (1986). 
\bibitem{222} R. Bousso and S. W. Hawking, {\it (Anti)-evaporation of Schwarzschild-de Sitter black holes}, Phys. Rev. {\bf D 57} (1998), 2436-2442.
\bibitem{GP} P. Ginsparg and M. J. Perry, Nucl. Phys. {\bf B222}, 245 (1983).
\bibitem{Barcelo} L. C. Barbado, C. Barceló and L. J. Garay, {\it Hawking radiation as perceived by different observers}, Class.Quant.Grav. {\bf 28} (2011) 125021; L.C. Barbado, C. Barcelo and L. J. Garay, {\it Hawking radiation as perceived by different observers: an analytic expression for the effective-temperature function}, Class.Quant.Grav. {\bf 29} (2012) 075013.   
\bibitem{Kempf} A. Kempf {\it Quantum Group Symmetric Fock Spaces with Bargmann-Fock representations}, Lett. Math. Phys. {\bf 26}:1-12 (1992).
\end{thebibliography}
\end{document}